\documentstyle[12pt,epsf]{article}
\begin{document}

\title{\bf Constraints on superdense preon stars and their
formation scenarios}

\author{J. E. Horvath\\
Instituto de Astronomia, Geof\'\i sica e Ciencias Atmosf\'ericas\\
Rua do
Mat\~ao 1226, 05508-900 S\~ao Paulo SP, Brazil\\
foton@astro.iag.usp.br\\}

\maketitle \pagestyle{empty} \vskip5mm

\noindent {\bf Abstract} We address in this work the general
features of a possible compact stars composed by elementary
fermions beyond the quark level. The {\it locus} of these
hypothetic objects in the mass-radius plane is constructed for the
maximum mass (minimum radius) of the sequence of models in terms
of a compositeness scale only, and in fact this approach applies
for any composite model postulating fermions at or beyond the
preon level. We point out a constraint on the preon mass arising
from the applicability of the General Relativity structure
equations, leading to the questioning of the hypothesis of light
preons if the preon scale is high, provided classical compact
objects are enforced. Some remarks on the existence of superdense
stars of astrophysical and primordial origin are made and
discussed.

\noindent {\it keywords: preons, compact stars, cosmology.}


\section{Introduction}

Relativistic astrophysics and particle physics seem to have merged
completely in modern science. Starting from the pioneer works, the
developed interplay between both disciplines is far-reaching and
long-lasting. As a prime example, the quest for the internal
composition of actual ``neutron'' stars, as discovered in 1967
(Hewish et al. 1967) is still ongoing after several decades of
research. Landau's (1932) insight preceded this discovery by more
than 30 years, and Baade and Zwicky (1934) were among the first to
postulate an actual astrophysical site to produce these objects,
namely type II supernova thought to arise after gravitational
collapse of a massive star. Later in the '60s the emergence of the
quark model (that is, the acknowledgement that nucleons are
composed of more elementary fermions, confined in bubbles of the
order $\sim \, 1 \, fm$ at low energies) raised the possibility of
having deconfined matter inside neutron stars (Ivanenko and
Kurdgelaidze 1969; Itoh 1970; Collins and Perry 1975) . A radical
version of this idea was put forward later, namely, that a form of
cold quark matter could be the true ground state of hadronic
interactions (Bodmer 1971; Terazawa 1979; Chin and Kerman 1979;
Witten 1984) and thus ``neutron'' stars should be rather giant
quark balls, properly renamed as {\it strange stars}. Stars may be
the only place in the universe for the latter to form.

In spite of its success, the well-known incompleteness of the
Standard Model, needing at least 19 parameters of unknown origin
to work, prompted unification ideas based on the gauge theory
concepts extended to higher energies. Following the historical
examples of the atomic nucleus and the subnuclear partons, the
idea of seeking another compositeness scale and construct the
known hadrons and leptons out of a small set of more elementary
components (termed {\it preons} hereafter) raised in the '70s.
Examples of this approach were discussed by Pati and Salam (1983)
and Pati (1989), among many others (see D'Souza and Kalman 1992
for a more complete account). This ``bottom-up'' strategy to
understand nature building blocks is now somewhat superseded by
Theories of Everything starting from a whole symbiosis of
particles and forces and trying to understand how to go down in
energy to match the observations and experiments accessible today.
However, and despite of the attractive features of TOEs, it is not
obvious that they can successfully explain the physics of the
Standard Model and beyond, disconnected of the natural unification
scale by many orders of magnitude in energy, and it is entirely
possible that unification schemes can be derived from them instead
of the Standard Model gauge structure (Pati 2006).

Several theoretical expectations to strike the next interesting
scale in physics rely on the existence of {\it supersymmetry},
relating known bosons and fermions, broken below some scale
$\Lambda_{SUSY}$. Supersymmetry (SUSY) is very important for TOEs
as well, and in fact many of the preon models required
supersymmetry to work (Pati 1989). It is expected that
supersymmetry can be observed in colliders at an energy scale of
around $\approx \, 1 \, TeV$. In fact, experimental limits to the
compositeness of electrons have shown (Sabetfakhri 2000) that
preons are not needed down to $\ell \, \sim \, 10^{-17} \, cm$, a
value consistent with some SUSY preon models, which should show up
well above the present accelerator energies. Based on all these
expectations, we shall always adopt $1 \, TeV$ as a minimal scale
for compositeness below, keeping in mind that a viable scheme
could require a much higher value.

The very existence of compositeness of quarks and leptons would
merit a consideration related to the compact star problem.
Recently Hansson and Sandin (2005) and Sandin (2005) have
discussed the general features of hypothetic preon stars,
including their possible role in avoiding black hole formation
(Hansson 2006). Given that almost nothing is known of detailed
preon physics (i.e. masses, interactions, etc.), we limit
ourselves to the most general features, which are quite
model-independent, and show below the room left in the mass-radius
plane and related issues. Finally we point out the serious
difficulties encountered to form preon stars in the contemporary
and primordial universe.

\section{How much room for preon stars?}

With just minimal assumptions about a fermionic character of
preons, their lightness (they should compose light known
particles, like neutrinos), and an assumed minimal compositeness
scale, it is still possible to address the features of putative
preon stars. As a suitable framework for this task, Narain,
Schaffner-Bielich and Mishustin (2006) recently analyzed the
general features of stars made of fermions by integrating a
dimensionless form of the Tolman-Oppenheimer-Volkoff equations for
scaled equations of state valid for free fermions and also for
model interactions. The masses of preons (Pati 1989; Hansson and
Sandin 2005), are expected to be light or zero, certainly much
smaller than the mass of the composites (which include the
neutrinos in most models, Pati and Salam 1983). This means that
most of the composite states masses are probably from
interactions. This is why a "bag model" approach has been employed
in Sandin (2006) to represent the main features of the preon
matter. Adapting the results of Narain, Schaffner-Bielich and
Mishustin (2006), we may write for the maximum mass and (minimum)
radius of the spherically symmetric models the general expressions

\begin{equation}
M_{max} \, = \, 2.16 \times 10^{2}  {1 \,
\over{\sqrt{\lambda/10^{5} TeV \, fm^{-3}}}} \, M_{\oplus}
\end{equation}

\begin{equation}
R_{min} \, = \, 8.5 \times 10^{2} {1 \, \over{\sqrt{\lambda/10^{5}
TeV \, fm^{-3}}}} \, cm
\end{equation}

where $\lambda$ is proportional to the energy density needed to
fit the electron mass in this bagged approach. The maximum mass
model sets the scale for the masses expected in nature, being the
most compact ones, even though the stellar sequences contain
larger and less massive stars. The work of Narain,
Schaffner-Bielich and Mishustin (2006) has also shown that the
integration of a dimensionless TOV equation allows a universal
description of the maximum mass-minimum radii relation, both
scaling as $\lambda^{-1/2}$, which reads

\begin{equation}
{M_{max} \over{M_{\oplus}}} \, = \, 0.22 {R_{min}\over{cm}}
\end{equation}

Another obvious constraint is that the stars can not become black
holes, that is to say, that no matter how important the
(repulsive) interactions other than the ``bag constant" are, the
compactness of the stellar models is bound from above. These
unaccounted repulsive interactions can presumably be parametrized
by an effective shift in $\lambda$, resulting in an increase of
the maximum mass. If causality is ignored (that is, $\rho \, = \,
constant$), it is well-known (Shapiro and Teukolsky 1983) that the
maximum compactness parameter satisfies $2GM_{max}/R_{min}c^{2} =
8/9$. For any {\it causal} equation of state, calculations by
Haensel, Lasota and Zdunik (1999) give instead a factor $0.7$ for
the r.h.s, allowing one to write down the boundary separating
compact stars from black holes, in the same units, as

\begin{equation}
{M_{max} \over{M_{\oplus}}} \, = \, 0.83 {R_{min}\over{cm}}
\end{equation}

The parameter space allowed for preon stars is shown in Fig. 1
under these assumptions.

As stated, a minimum compositeness lengthscale $\ell \, \sim \,
10^{-17} \, cm$ has been assumed, according to the present
experimental limits. The preon ``stars" must necessarily be
car-sized objects or smaller, as already noticed by Hansson and
Sandin 2005. This region encompasses {\it any} stellar model made
from composite fermions, and the whole hierarchy of composed
objects down to a minimal scale discussed in Hansson (2006).

It is clear from these considerations that allowed region can
extend all the way down to {\it microscopic} values of $M_{max}$
and $R_{min}$. However, the structure calculations do not make
sense unless the pressure, density and other quantities can be
defined as classical quantities. This would be possible in turn if
preons are not too light, because otherwise their Compton
wavelength $\lambda_C$ would be bigger than the radius of the
''star" $R_{min}$. From this condition, $\lambda_C \, \ll \,
R_{min}$ we obtain the following bound

\begin{equation}\label{}
m_{p} \, \gg \, 2 \times 10^{-5} {\sqrt{\lambda/10^{5} TeV \,
fm^{-3}}} \, eV .
\end{equation}

For the adopted minimum scale $\lambda$ we find that the preon
mass $m_{p}$ starts to conflict with the determinations (Goobar et
al. 2006) of the neutrino mass $m_{\nu} \leq \, 0.3 \, eV$, for a
very high energy density $\lambda \, \sim 10^{13} TeV fm^{-3}$),
or $\lambda^{1/4}$ around $10^{6}$ times the natural QCD bag scale
$B^{1/4} \, \sim \, 150 \, MeV$. It is not presently known whether
this scale is too high for the next level of compositeness. The
compact star description would be correct below this scale.
However, if the compositeness produces a value above $\sim 10^{13}
TeV fm^{-3}$, we may still employ it if we abandon the idea that
preons must be lighter than their bound states. On the other hand,
we do not have any reliable description for preon stars if they
should be considered as quantum objects.

\section{Formation of preon stars and preon nuggets}

The mismatch between the maximum compact star mass made of preons
$\sim \, 100 \, M_{\oplus}$ and the neutron (or quark) scale can
not be overemphasized: the latter is found to lie around $1 \,
M_{\odot}$ because the hadronic scale is controlled by $B \sim \,
60 \, MeV fm^{-3}$, whereas the correspondingly larger $\lambda$
energy density gives rise to a factor $\surd (B_{QCD}/\lambda \,
\sim 10^{-4}$, leading to earth-like masses for the preon stars.
Therefore, if an actual hadron (or quark) star is made to
collapse, for example, being pushed over its maximum mass by
accretion from a companion, the scale at which a preon region can
form at the center must be much smaller than the Schwarzschild
radius of the whole collapsing object. The falling of a mass $\gg
\, M_{max}$ onto the center suggests that preon stars, with an
average static density $<\rho> = 3.7 \times 10^{26} \,
(R_{min}/cm)^{-2}$ can not form in the contemporary universe,
rather becoming black holes of stellar mass size. A solution would
be to couple preons to some kind of massless particle that can be
extremely efficiently radiated during collapse, as already pointed
out by Hansson and Sandin (2005). This would keep the
Schwarzschild radius smaller than the actual radius of the
collapsing configuration, thus avoiding the formation of a black
hole, but its realization remains to be convincingly demonstrated.

An alternative would be to form the preon objects when they bind
forming quarks and leptons, in the early universe, hereafter named
{\it preon nuggets}. They could cool and become self-gravitating
eventually, leaving light preon stars, provided they form and
survive. The naive temperature at which this happens can be found
by assuming a radiation gas for the preon matter and demanding its
pressure to be positive (that is, its kinetic term should remain
larger than the energy density associated with $\lambda$). Using
the same values as above, this estimate yields

\begin{equation} T_{p} \, \sim \, 12 \times (10/{g_p})^{1/4} \, GeV
\end{equation}

where $g_{p} = n_{b} +(7/8)n_{f}$ counts the number of
relativistic bosonic and fermionic degrees of freedom $n_{b}$ and
$n_{f}$. A preon to quark-lepton phase transition seems to require
$g_{p} > g_{q-l}$ as well (Nishimura and Hayashi 1987), a
condition which is not realized for the most economic description
(Harari 1979) of substructure, but posses no problems for more
elaborated schemes (Harari and Seiberg 1981). Note that all quark
flavors, with exception of the top, and all lepton types would
contribute to $g_{q-l}$ at a temperature $\sim \, T_{p}$, thus
tightening the bound $g_{p} > g_{q-l}$ which is non-trivial and
must be studied on a case-by-case basis.

The horizon at the onset of the quark-lepton era is $H^{-1} \sim
\, 3 \times 10^{2} ({g_{p}}/10)^{3/2} \, cm$, a factor of 3 below
the radius of the maximum mass model. Thus only preon structures
around $\leq \, 1 \, M_{\oplus}$ could be formed, if at all, in
the cosmological setting (Hansson and Sandin 2005). This should be
considered as an absolute upper limit for primordial objects
originated in the limits on compositeness as discussed above.

Given that $m_{p} \ll m_{l}$, with $m_{l}$ any lepton mass, and
quite possibly zero, we may think analogously to the QCD phase
transition and ask whether the preons would like to stay as such
(forming ``preon nuggets"), doubling the celebrated ``Witten
effect" (Witten 1984) at higher energies. In such a scenario, when
the cosmological temperature falls below $T_{p}$, the universe
remains mostly in the preon phase with some supercooling, until
nucleation of quarks and leptons proceeds through bubble
nucleation (first order transition). Quark-lepton bubbles release
latent heat and expand, heating the surrounding preon matter,
until the latter pushed into small regions can provide enough
pressure to stop its contraction. The number of particles trapped
in nuggets would depend on the degree of supercooling and
transport properties in the process. However, analogously to the
discussed strange quark nugget scenario (and technicolor matter as
well, Frieman and Giudice 1991) it is important that nuggets can
loss energy without necessarily loosing the conserved charge that
stabilizes them as non-topological solitons, as baryon number or
technibaryon number are. It is not clear whether a preon model
possessing such a conserved charged (e.g. preon number) can be
constructed. For instance, in ordinary QCD global symmetries like
baryon number or isospin can not be broken (Vafa and Witten 1984),
and also that massless states do not form from massive
constituents. These restrictions do not apply, for example, to
massless preon theories with gauge Yukawa interactions as
discussed in Pati (1989). The point here is that the formation or
absence of preon nuggets depends on the specific model assumptions
(see Das and Laperashvili 2006; Dugne, Fredriksson and Hansson
2002 and Burdyuzha et al. 1999 for very recent models) to a point
which is not possible to state anything firm today. However, a
large class of models are ruled out from the scratch,
independently of other conditions.

In any case, if formed, these concentrations should be fragile
against evaporation into quarks and hadrons at intermediate
temperatures. The gravitational potential is initially
unimportant, and does not help much either because of the limit
given by the horizon. Clearly, the situation is much worse if the
compositeness scale happens to be higher. This arguments cast
doubts on the very formation of preon nuggets, but in any case,
this issue remains to be thoroughly investigated.

\section{Conclusions}

Are there ``preon stars" in the present universe? The answer is
not simple, but the payoff potentially large. We have discussed
the region in the mass-radius plane available in a one-parameter
approach (just an energy density $\lambda$) and found results
consistent with more detailed calculations (Hansson and Sandin
2005) suggesting robust predictions for them. An important clue
for the problem of preon masses related to these objects has been
pointed out, namely the validity of a classical description
suggesting a lower limit on $m_{p}$. The formation of these
superdense objects, separated by ``ordinary" neutron/quark stars
by a jump of 12 orders of magnitude in the average density, is
problematic not only in contemporary scenarios but in the early
universe as well. These arguments should be further reexamined
before a reasonable answer to the existence question can be given.

\section{Acknowledgments:}

This work was supported by Funda\c c\~ao de Amparo \`a Pesquisa do
Estado de S\~ao Paulo and the CNPq Agency (Brazil).

\vfill\eject

\bigskip
\noindent Fig. 1. The {\it locus} of preon stars (and higher level
of compositeness fermion objects) in the M-R plane. The allowed
trapezoidal region is limited on the right by the minimum
compositeness scale, on the bottom by the maximum mass- minimum
radius given by eq.(3) and atop by the black hole boundary eq.(4).
Two possible sequences of models are sketched, both without
interactions (solid line) and with some repulsive interaction
(dashed line) driving the models closer to the black hole limit.

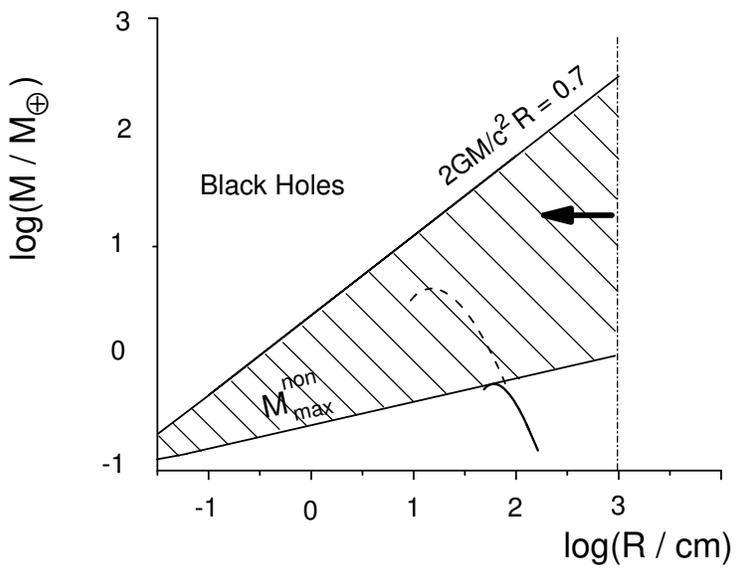
\end{document}